\documentclass[12pt]{article}

\usepackage{epsfig}
\usepackage{amsbsy}
\usepackage{amsmath}

\begin{document}

\title{Moments of the particle phase-space density at freeze-out and coincidence probabilities}
\author{A.Bialas, W.Czy\.{z} and K.Zalewski
\\ M.Smoluchowski Institute of Physics
\\ Jagellonian University, Cracow\footnote{Address: Reymonta 4, 30 059 Krakow,
Poland, e-mail: bialas@th.if.uj.edu.pl, zalewski@th.if.uj.edu.pl}
\\ and\\ Institute of Nuclear Physics, Cracow}
\maketitle

\begin{abstract}
It is pointed out that the moments of phase-space particle density at
freeze-out can be determined from the coincidence probabilities of the
events observed in multiparticle production. A method to measure the
coincidence probabilities is described and its validity examined.

\end{abstract}

 \noindent PACS numbers 25.75.Gz, 13.65.+i \\Bose-Einstein
correlations, interaction region determination.
 \vspace{1cm}

Recently we have proposed \cite{BCZ} a method for measuring the average of the
particle density in multiparticle phase space. Generalizing the approach used
by Bertsch \cite{BER} for single particle phase space densities, we defined the
average density for a final state consisting of $M$ particles as

\begin{equation}\label{}
\langle D_M \rangle = M \int\!\!dXdK\;W^2(\textbf{K},\textbf{X}).
\end{equation}
Here $W(\textbf{K},\textbf{X})$ is the $M$-particle distribution
 function and the
integrations are over the $3M$ components of the  position vectors and
over the $3M$ components of the  momentum vectors of the particles.
Following Bertsch we replaced the emission function depending on the
four-vectors $K$ and $X$ by a  function depending on the three-vectors
$\textbf{K}$ and $\textbf{X}$ at some representative time $t_0$ which does not
need to be specified. This density distribution represents the situation at
freeze-out.

In the present paper the method of measurement is generalized to
 arbitrary integer moments $(l)$ of the phase space density

\begin{equation}\label{defden}
\langle D_M^l \rangle = M^l \int\!\!dXdK\;W^{l+1}(\textbf{K},\textbf{X}).
\end{equation}
As our argument follows closely that of \cite{BCZ}, we shall be very brief. For details the reader can consult \cite{BCZ}.

The method uses the coincidence probabilities defined as
follows. An $M$-particle final state could be characterized by specifying the
momenta of the $M$ particles $\textbf{p}_1,\ldots,\textbf{p}_M$. Then, however,
no two observed events would be identical. It has been proposed instead
\cite{BCW}-\cite{BIC4}, to introduce
momentum bins of finite width and to consider two events as identical, if they
have the same distribution of particles among the bins. Let us denote the total
number of events by $N$ and the number of states which occurred $l$ times by
$N_l$. If a state occurs $k>l$ times it contributes to $N_l$ its weight $\left(
\begin{array}{c}
    k \\
    l \
  \end{array}\right)$. Then
the number of $l$-fold coincidences is defined by

\begin{equation}\label{1}
  C_M^{exp}(l) =\left( \begin{array}{c}
    N \\
    l \
  \end{array}\right)^{-1}N_l.
\end{equation}
$C_M^{exp}(l)$ is the ratio of the number of sets of $l$ events with equivalent
final states to the total number of sets of $l$ events. Thus, for large $N$, it
is the probability of an $l$-fold coincidence. The special case $l=2$ had been
considered, in a different context, by Ma \cite{MA1}. The
coincidence probability depends, of course, on the binning. For a given binning
it is in principle not difficult to find from a data sample the coincidence
probabilities. In practice the limiting factor is statistics.

We will now show that for suitably defined bins the {\it effective} coincidence probabilities

\begin{equation}\label{ceffec}
\hat{C}_M(l+1) \equiv   \left(\frac{(2\pi)^{3M}}{M}\right)^l\langle D_M^l \rangle
\end{equation}
 can be approximated by $C_M^{exp}(l+1)$\footnote{ The power of
$2\pi$ in (\ref{ceffec}) appears because for $M$ free particles there is one quantum state per
volume $(2\pi)^{3M}$ in phase space.}.

To see that, we first express the measured l-fold coincidences (\ref{1})  by the $3M$-dimensional distribution of
momenta

\begin{equation}\label{}
  w(\textbf{K}) = \int\!\!dX\;W(\textbf{K},\textbf{X}).
\end{equation}
Let us denote the $3M$-dimensional momentum bins by $j = 1,\ldots ,J$ and their
volumes by $\omega_j$. Then the probability that  an event
corresponds to a point in bin $j$ is

\begin{equation}\label{}
  P_j = \int_{\omega_j}\!\!dK\;w(\textbf{K})
\end{equation}
and the average density in bin $j$

\begin{equation}\label{}
  w_j = \frac{P_j}{\omega_j}.
\end{equation}
The probability that $l$ events chosen at random correspond each
to a point in bin $j$ is $P_j^l$.
Therefore, the probability that $l$ events chosen at random are  the same is

\begin{equation}\label{cexper}
  C_M^{exp}(l) = \sum_j P_j^l = \sum_j (\omega_j)^l (w_j)^l.
\end{equation}

Next, we observe that, as seen from  (\ref{defden}),

\begin{equation}\label{}
\langle D^l \rangle = M^l \sum_j \int_{\omega_j}
\!\!dK\int\!\!dX\;W^{l+1}(\textbf{K},\textbf{X}).
\end{equation}

In order to proceed further, following \cite{BCZ}, we restrict ourselves to the phase-space distributions of the general form

\begin{equation}\label{assump}
  W(\textbf{K},\textbf{X}) = \frac{1}{(L_xL_yL_z)^M}G(\textbf{X}/L)w(\textbf{K}).
\end{equation}
Here $\textbf{X}/L$ is a $3M$-dimensional vector with components $(X_1 -
\overline{X}_1)/L_x,\ldots, (Z_1 - \overline{Z}_M)/L_z$. The parameters $L_x,
L_y, L_z, \overline{X}_1,\ldots,\overline{Z}_M$ are in general functions of
$\textbf{K}$. They could also be different for different kinds of particles.
Function $G(u)$ satisfies the conditions

\begin{equation}\label{}
  \int\!\!du\;G(u) = 1;\quad \int\!\!du\;uG(u) = 0;\quad \int\!\!du\;u^2G(u) =
  1,
\end{equation}
where all the integrations are $3M$-fold. These relations imply

\begin{eqnarray}\label{}
  \int\!\!dX\;G(\textbf{X}/L) &=& (L_xL_yL_z)^M;\\
  \langle X_1 \rangle &=& \overline{X}_1,\ldots, \langle Z_M \rangle =
  \overline{Z}_M;\\
  \sigma^2(X_i) = L_x^2;\quad \sigma^2(Y_i) &=& L_y^2;\quad \sigma^2(Z_i) =
  L_z^2;\quad \mbox{for $i=1,\ldots,M$}.
\end{eqnarray}
The first condition ensures that $w(\textbf{K})$ is the observed
$3M$-dimensional momentum distribution. The other two yield the physical
interpretation of the parameters
$\overline{X}_i,\overline{Y}_i,\overline{Z}_i,L_x,L_y,L_z$.

Using the Ansatz (\ref{assump}) and the definition (\ref{ceffec}) one finds

\begin{equation}\label{}
 \hat{C}_M(l+1) =(2\pi g_{l+1})^{3Ml}
 \sum_j   \left(\prod_{m=1}^M(L_xL_yL_z)_j^{(m)}\right)^{-l}
\int_{\omega_j}
\!\!dK\;w^{l+1}(\textbf{K}).
\end{equation}
with $(g_{l+1})^{3Ml}\equiv \int\!\!du\;G^{l+1}(u)$.

Comparing this formula with formula (\ref{cexper}) it is seen that
if the bin sizes are

\begin{equation}\label{omo}
  \omega_j =
\prod_{m=1}^M\frac{(2\pi g_l)^3}{(L_xL_yL_z)_j^{(m)}}
\end{equation}
we obtain

\begin{eqnarray}
\hat{C}_M(l) =
C^{exp}_M(l)\frac{\sum_j\frac1{\omega_j}\int_{\omega_j}dK[w(\textbf{K})]^l}
{\sum_j[\frac1{\omega_j}\int_{\omega_j}dKw(\textbf{K})]^l}.
\end{eqnarray}

Eq. (\ref{omo}) is rather general and can be applied to an entirely arbitrary
discretization procedure. In the simple (but probably most practical) case when
$L_xL_yL_z$ does not depend on $\textbf{K}$, one can choose bins of constant
lengths $\Delta_x,\Delta_y,\Delta_z$ along the axes
 $K_x,K_y,K_z$, respectively. Then the condition for
the size of the bin is

\begin{equation} \label{}
\Delta_x\Delta_y\Delta_z= \frac{(2\pi g_l)^3}{L_xL_yL_z}.
\end{equation}
and $\omega_j=[\Delta_x\Delta_y\Delta_z]^M$.

These results, being very similar to that of \cite{BCZ}, invite several
comments which were elaborated there at length. Here we only repeat the
main points.

As a first approximation we may omit the ratio of averages in the square
bracket, giving $\hat{C}_M(l) =C^{exp}_M(l)$. This approximation
becomes exact when function $w(\textbf{K})$ is constant within each bin $j$.
The condition is on the volume of the bin, but does not constrain its
shape. This freedom can be used to improve the approximation. Bins
should be chosen narrow in directions where function $w(\textbf{K})$
changes rapidly and broad in directions with little or no variation of
$w(\textbf{K})$. For given variability of $w(\textbf{K})$ our
approximation improves when the bins become smaller, i.e. when the
product $L_xL_yL_z$ increases. Thus we expect the method to work much
better for central heavy ion collisions than e.g. for
$e^+e^-$annihilations. With very good statistics, or a reliable Monte
Carlo event generator, one could estimate the term, which here we
replaced by unity, and thus improve the approximation. In order to find
the bin size it is necessary to know the volume in coordinate space
$(L_xL_yL_z)$ and the integral $\int\!\!du\;G^l(u)$. Information about
the volume may be provided by interferometric measurements. The integral
is not very sensitive to the shape of the distribution. For a
rectangular box $2\pi g_{l}= \pi/\sqrt{3}$. For Gaussians $2\pi g_l=
\sqrt{2\pi}/[\sqrt{l}]^{1/(l-1)}$.

In conclusion, we have generalized the results of \cite{BCZ} to
measurements of the higher moments of the phase-space density of
particles produced in high-energy collisions. It turns out that these
moments can be approximated by the measured coincidence probabilities of
the appropriately discretized events. The optimal discretization is
suggested and shown to depend critically on the volume of the system in
configuration space. Thus the actual measurements would require this
additional information to be effective.

\vspace{0.3cm}
{\bf Acknowledgements}
\vspace{0.3cm}

Discussions with Robi Peschanski and Jacek Wosiek are highly
appreciated.

\end{document}